\documentclass{eas}
\usepackage{graphicx}
\newcommand{\der}{{\rm d}}

\newcommand{\be}{\begin{equation}}
\newcommand{\ee}{\end{equation}}
\newcommand{\ba}{\begin{eqnarray}}
\newcommand{\ea}{\end{eqnarray}}
\newcommand{\brr}{\begin{array}}

\newcommand{\err}{\end{array}}

\begin{document}

\TitreGlobal{Mass Profiles and Shapes of Cosmological Structures}

\title{caustics in dark matter haloes}
\author{Roya Mohayaee}\address{IAP, Institut d'Astrophysique de Paris, 98 bis
bd Arago, Paris, France}
\author{Stephane Colombi$^1$}
\author{Bernard Fort$^1$}
\author{Raphael Gavazzi}\address{ Laboratoire d'Astrophysique, OMP, 
14 Av Edouard Belin, F-31400 Toulouse, France}
\author{Sergei Shandarin}\address{Department of Physics and Astronomy, University 
of Kansas, KS 66045, U.S.A. }
\author{Jihad Touma}\address{Department of Physics, 
Center for Advanced Mathematical Sciences, 
American University of Beirut, Beirut, Lebanon
}
\runningtitle{Caustics in Dark Matter Haloes}
\setcounter{page}{23}
\index{Roya Mohayaee, A.}
\index{Stephane Colombi, B.}
\index{Bernard Fort, C.}
\index{Raphael Gavazzi, D.}
\index{Sergei Shandarin, E.}
\index{Jihad Touma, F.}
%
\begin{abstract}

Caustics are formally singular structures, with infinite density, that form in 
collisionless media. The non-negligible velocity dispersion of dark matter
particles renders their density finite. We evaluate the maximum density of the caustics
within the framework of secondary infall model of formation of dark matter haloes.
The result is then used to demonstrate that caustics can be probed by properly stacking the
weak-lensing signal of about 600 haloes. CFHTLS accompanied by X-ray observations and the
space-based experiments like SNAP or DUNE can provide us with the required statistics and
hence a way of distinguishing between the viable dark matter particle candidates. The
extension of our results to more realistic models including the effects of
mergers of haloes is briefly outlined.

\end{abstract}

\maketitle
%
\section{Introduction}

Under the gravitational instability, cold streams of dark matter with different
velocities cross at {\it caustic} surfaces. The density of dark matter at
caustics formally diverges and in the regions bounded by caustics the velocity
field is multi-valued. 
The full description of the evolution of the fluid is provided by the
Vlassov-Poisson equation, in 6-dimensional 
phase space. Direct numerical integration of
this equation remains a challenging 
task and so far is best achieved for systems with 
low number of dimensions and or 
with strong symmetry constraints (see {\it e.g.} Alard \& Colombi
2004 and references therein). The most popular 
alternative to the full integration 
of this equation is to sample the phase space with discrete particles.
However, extremely large number of particles is needed 
to sample the fine-structure of the
phase space and significant softening of the forces is required to avoid
spurious collisional effects which would smear out 
the caustics (see {\it e.g.} Binney 2004, Melott et al. 1997).

Until a full 3-dimensional solution of Vlassov-Poisson equation or 
ground-breaking resolution in N-body simulations or other
innovative solutions are achieved, 
(semi-)analytic modeling of haloes can provide approximative description of the
dynamics at fine scales.

\begin{figure}[h]
\centering
\includegraphics[width=10cm]{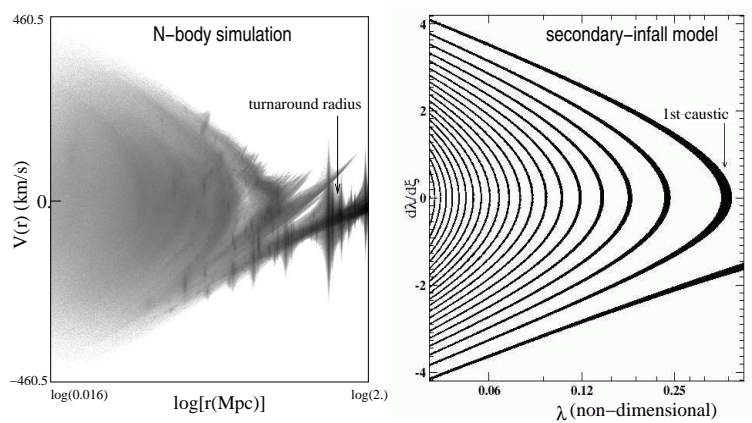}
\caption{{\it Left panel}:
Phase space diagram (radial velocity versus radius) of a Milky-Way-type 
halo, taken from a PM simulation with $16$ millions of particles 
on a $1024$ grid in a $4$ Mpc simulation box 
realized at IDRIS computing center (CNRS).
The halo has about 8 million particles in its virial radius
and contains only 40 substructures, as they are rapidly disrupted due 
to softening of the forces in the PM routine.
The smoothness of the gravitational potential
(there are many particles per softening
length) enables us to see the caustics since the phase-space 
structure can survive the relaxation effects.
{\it Right panel}: Secondary infall model's prediction of the phase space of 
a spherical halo in smooth accretion. 
The caustics in the model are far better
resolved as compared to the left panel, due to the simple assumption of
sphericity, no abrupt major merger and no substructures.
The panel clearly shows that the streams {\it cool down} as they collapse: 
a consequence of phase-space volume 
conservation as filamentation builds up.
}
\label{singlehalo}
\end{figure}

Analytic works on the halo
density profile, and its caustics, started with the works of
Gott (1975) and Gunn (1977).
In an Einstein-de Sitter Universe a spherical over-density
expands and then turns around to collapse.
After collapse and at late times, the fluid motion becomes
selfsimilar: its form
remains unchanged when length are re-scaled in terms of the radius of the 
shell which is currently turning around ($r_{\rm ta}$) and falling onto the galaxy.
Self-similar solutions give power-law density profiles which is convolved with
many small-scale spikes
{\it i.e.} caustics
(Fillmore and Goldreich 1984, Bertschinger 1985a, 1985b).

In the self-similar accretion model, Newton's law
is reformulated as
\be
{d^2\lambda\over d\xi ^2}+{7\over 9}{d\lambda\over d\xi}
-{8\over 81}\lambda=-{2\over 9\lambda^2} M(\lambda)\,,
\label{newtonnondimensional}
\ee
by using the non-dimensional variables
$$
\lambda = {r(t)\over r_{\rm ta}(t)}\quad;\quad
\xi={\rm ln}\left({t\over t_{\rm ta}}\right)\quad;\quad
M(\lambda)={3\over 4\pi}{m(r,t)\over \rho_H r_{\rm ta}^3}\,,
\label{nondim1}
$$
where $m(r,t)$ is the mass inside a radius $r$ [$m(r,t)$ is not
constant due to shell-crossing],
$\rho_H=1/6\pi Gt^2$ 
and $t_{\rm ta}$ is the turnaround time for
a given particle ({\it i.e.} when the particle is at its largest radius). 
This reformulation also assumes the power-law initial condition
\be
{\delta M_i\over M_i}=\left({M_i\over M_0}\right)^{-\epsilon}
\label{eq:initialmass}
\ee
where $M_0$ is some reference mass and $M_i$ is 
the unperturbed mass within the initial radius $r_i$,  
It might appear that the asymptotic self-similarity 
is a direct outcome of the power-law initial mass profile. 
However, it has been shown that starting from a generic initial condition, 
a power-law density profile is achieved
prior to the first turnaround time (Moutarde et al 1991). The turnaround radius 
increases with time as
$
r_{\rm ta} \approx t^{(2/3)(1+1/3\epsilon)}
$ and hence the mass inside the turnaround radius increases as
$M_{\rm ta}\approx (1+z)^{-1/\epsilon}$. This scaling compares with that
of the characteristic mass in a scale-free hierarchical 
clustering Universe $M_*\approx (1+z)^{-6/(n+3)}$
where the power-spectrum of initial density 
fluctuations is $P(k)\approx k^n$. 
A central point mass perturbation corresponds to
$\epsilon=1, r_{\rm ta}\approx t^{8/9} ; M_{\rm ta}\approx 1/(1+z)$ which is
the only case considered here. Extension of our results to more realistic
initial conditions ({\it e.g.} $\epsilon\sim 0.2$) will follow in forthcoming works.

For a perfectly cold dark matter medium, the density profile close
to the $k$th caustic at $\lambda_k $ is (Bertschinger 85b)
\begin{equation}\label{eq:rho3Dsing}
  \frac{\rho_0(\lambda)}{\rho_H}= \left(\frac{\pi^2}
{4\sqrt{-2\lambda_k''\,}}\,\frac{e^{-2 \xi_k/3}}{\lambda_k^2 }\right)
\frac{1}{\sqrt{\lambda_k-\lambda}}\,;
\qquad\qquad \sigma=0.
\end{equation}
(The values of the various quantities $\xi_k$, $\lambda_k$, $\lambda_k''$
are given in Bertschinger 1985b.)

When the temperature of particles is not strictly zero,
caustic positions are shifted by a small value $\delta\lambda$ and
the density near the caustic is modified as:
$
\rho_\sigma(\lambda)= \int \der v\, \rho_0\,[\lambda - \delta\lambda(v)]\,
f(v)\,.
$
A simple top-hat velocity distribution function, $f(v)$, yields
\begin{equation}\label{eq:rho3Ddef}
\frac{\rho_\sigma(\lambda)}{\rho_H} = 
\left(\frac{\pi^2}
{4\sqrt{-2\lambda_k''\,}}\,\frac{e^{-2 \xi_k/3}}{\Delta_k\lambda_k^2 }\right)
\left\{
\begin{array}{ll}
\sqrt{\lambda_k^+-\lambda} - \sqrt{\lambda_k^--\lambda} & 
\quad{\rm for}\quad\lambda< \lambda_k^-\;,\\
\sqrt{\lambda_k^+-\lambda} & \quad{\rm for}\quad\lambda_k^-<\lambda<\lambda_k^+\;, \\
0 &\quad{\rm for}\quad\lambda> \lambda_k^+\;,
\end{array}\right.
\end{equation}
for the density near the $k$-th caustic,
where $\lambda_k^-=\lambda_k-\Delta_k$ and $\lambda_k^+=\lambda_k+\Delta_k$
and the thickness of the $k$-th caustic is $\Delta_k$.  
A simple re-scaling demonstrates that these densities are universal: they are
valid for all caustics and irrespective of size and mass of their halo.
(for full details see Mohayaee \& Shandarin 2005).

\begin{figure}[h]
\centering
\includegraphics[width=4cm]{mohayaee_fig2.eps}\qquad
\includegraphics[width=5.75cm]{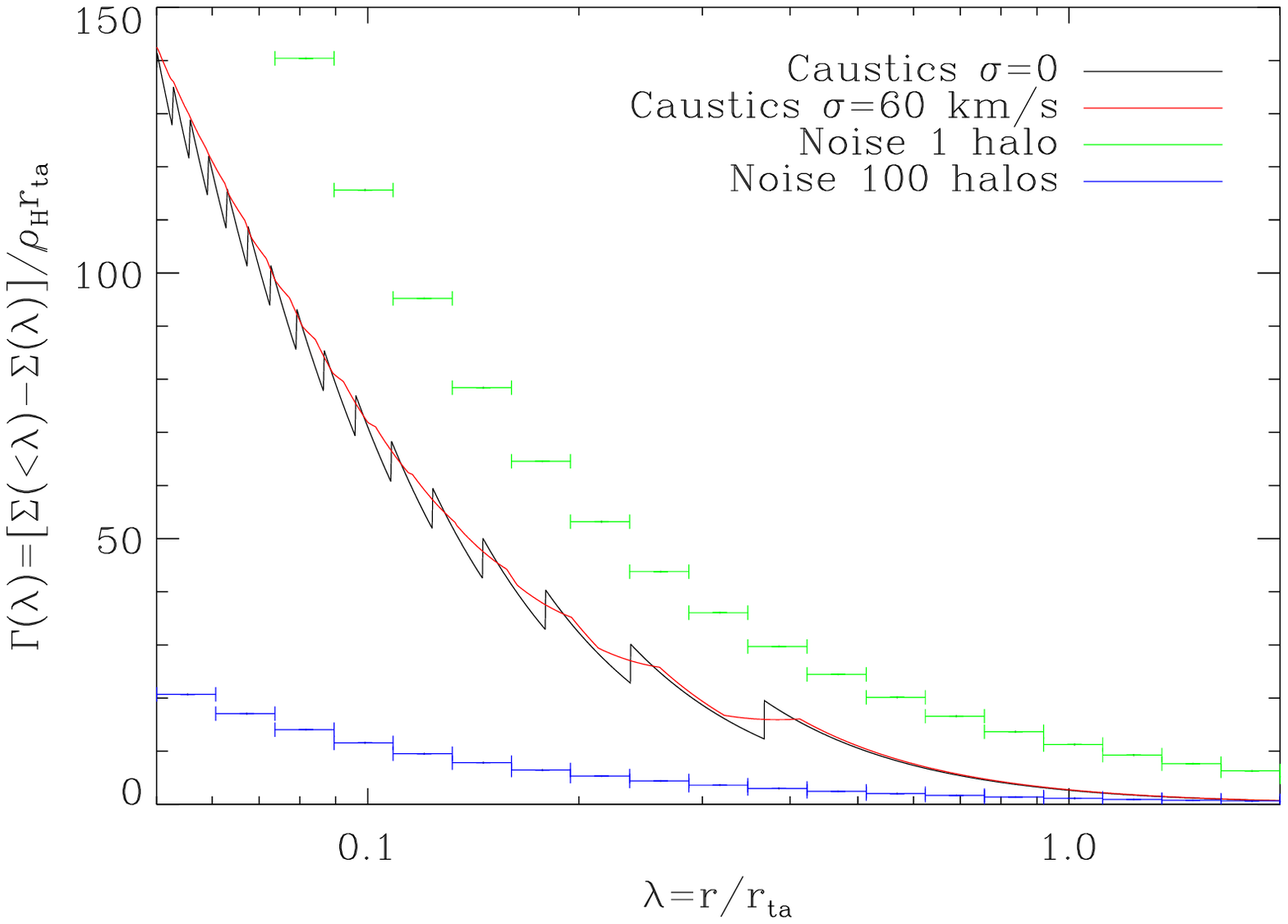}
\caption{ 
{\it Left panel} 
{\it Left panel ; Main plot}: The top line (filled circles) 
shows the separation of caustics, 
$\lambda_k-\lambda_{k+1}$, as a function of the distance, $\lambda_k$, 
from the center. The bottom line
shows the thickness of the caustics.
{\it Left panel ; Inset:} The ratio of the thicknesses of 
the caustics to their separations , 
$2\delta\lambda_k/(\lambda_k-\lambda_{k+1})$, is shown as a function of 
their radii.
This panel demonstrates that in the course of the gravitational evolution
streams remain well-isolated from each other in spite of the fact that
their separations diminish.
{\it Right Panel}:
Pseudo-shear is plotted for two
    values of $\sigma$: cold medium $\sigma=0~$ (black curve) and warm medium
    $\sigma\sim 60~$ km/s (red curve). Unphysically large value of velocity
dispersion is therefore required to smear out the caustics.
The green (resp. blue) binned curve is the
    noise level for one (resp. 100 stacked) halo(es).
}
\label{fig:gamma}
\end{figure}

Caustics could have immediate impact for dark matter search experiments. 
Their high density and the fact that they 
stay well-seperated from each other (see left panel of Fig \ref{fig:gamma})
can leave detectable fluctuations in the annihilation
products of the dark matter particles [if it constitutes 
for example of self-annihilating axions or
neutralinos] (see {\it e.g.} Sikivie \&  Ipser 1992, Sikivie et al. 1997) and 
or in the weak lensing data, which we shall study here 
[see {\it i.e.} Hogan 2001, Charmousis 2003, Gavazzi et al. 2005, Onemli 2004
and 2005].

For weak lensing, the Abel integral relates the 3-dimensional density ($\rho$) 
and the 2-dimensional density profiles ($\Sigma$) by (see Gavazzi et al. 2005
and references therein for full details)
\begin{equation}
  \Sigma(\lambda)= 2 r_{ta} \int_{\lambda}^\infty \frac{\rho(\lambda') 
\lambda' \der \lambda'}{\sqrt{\lambda'^2-\lambda^2}}\;.
\end{equation}
We numerically integrate the above expression, for the density profiles
(\ref{eq:rho3Dsing}) and (\ref{eq:rho3Ddef}).
We define a pseudo-shear:
$\Gamma(\lambda)=(\overline{\Sigma}-\Sigma)/\rho_H r_{ta}$
and the corresponding noise level $\Gamma_N$.
For an EdS cosmology, and considering an annulus of inner and outer radii
$\lambda_1$ and $\lambda_2$ respectively, it is straightforward to write
$\Gamma_N$ in the following units:
\be
  \Gamma_N(\lambda_1,\lambda_2) = \, 2.16\, \frac{D_{os}}{D_{ls}} 
\left({5 {\rm Mpc}\over r_{ta}} \right)^2 (1+z_l)^3 \times 
 \sqrt{{30\, \mathrm{\rm arcmin}^{-2}\over n}} 
  \left({\sigma_e\over 0.3}\right) {1\over \sqrt{\lambda_2^2-\lambda_1^2}}\;.
\ee

Fig. \ref{fig:gamma} shows pseudo-shear as a function of distance from the
center of the cluster for 
the thermal velocity dispersions $\sigma=0~{\rm km/s}$ and
$\sigma=60~{\rm km/s}$.
Comparing these curves, one can see that the sawtooth patterns due to 
caustics survive significantly high temperatures.
Next, we consider the noise level for a fiducial halo at redshift $z_l=0.3$
[since intermediate redshift haloes at $z_l\sim 0.2-0.5$ are the most likely
targets; see Gavazzi et al. 2005 for further details]. 
and a turnaround radius $r_{ta}=5{\rm Mpc}$ which is a typical value for
clusters (upper green binned curve). With a single halo 
the detection of caustics
is impossible. Since, the noise falls with the square-root of the number of
lenses (as well as the source), if one could stack 
the signal from a few tens of
clusters, the noise level will be low enough to
be sensitive to caustics as a 
whole (lower blue curve). Systematic error can 
arise due to the intrinsic
ellipticity of the clusters. The precise effect of which can be 
well-studied with the help of N-body simulations in future works. However,
due to its significant separation from the second caustic and the turnaround
radius, it is
likely that the first outer caustic would survive smearing due to
substructures and ellipticity of the haloes.

\begin{figure}[h]
   \centering
\includegraphics[width=10.5cm]{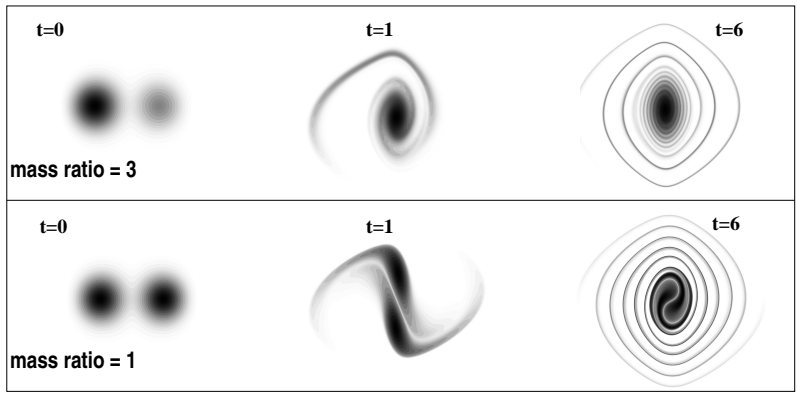}
\caption{
Phase space of a 1D-1D simulation of merger of two haloes
using the waterbag code of Colombi and Touma (2005). 
The simulation indicates that
in the merger of two haloes, phase-space folds do not mix and 
the topology of caustics remains intact.
}     
\label{fig:twohaloes}
\end{figure}

Although in this work we have considered dynamics of a simple isolated haloes,
we believe that the results can be extended to haloes growing 
in cosmological environment.
One of the short-comings of the secondary infall model could be that it ignore
the hierarchical scenario of merger. Merger might lead to distorted topology for the
caustics and hence render them practically non-detectable.
However, minor mergers which are on average isotropic are well-described by the
self-similar model, which assumes a spherically symmetric accretion. Thus such
mergers would not erase the caustics.  
Indeed even if the condition of isotropy is not satisfied, the caustics retain
their shapes quite well. The numerical solution to a 
1-dimensional Vlassov-Poission equation clearly suggests 
this fact in Fig. \ref{fig:twohaloes}. 
In mergers of unequal mass haloes the smaller halo wraps around the
larger halo without disrupting
its phase-space structure. In fact it populates it in a coherent manner as
shown in the top panel of Fig. \ref{fig:twohaloes}.

In a major merger, a similar situation follows. The phase sheets of the two
caustics do not mix, However, they fold closely together in phase
space [bottom row of Fig. \ref{fig:twohaloes}].
Thus, Fig \ref{fig:twohaloes} demonstrates that in the merger of two haloes
the topology of caustics remains intact.

Another important feature demonstrated well in Fig. \ref{fig:twohaloes} is the
gravitational cooling effect. With the increase of dynamical time ($t$) the
coarse-grain velocity dispersion which is related to the number of folds in the phase
space increases. However, each fold cools down and becomes thiner, a
surprising feature of a system with negative specific heat, but which can be
understood roughly by the conservation of phase-space volume.

For a perfectly cold dark matter with zero velocity dispersion, caustics 
are singular objects with zero thickness and infinite density which form
inevitably in the course of gravitational collapse. 
The small velocity dispersion of dark matter particles renders 
the density and the thickness of the caustics finite. The latter quantities,
if observed, could provide direct evidence for the existence of dark matter
and put bounds on its mass.

We have shown that the existence of dark matter caustics could be probed
by properly stacking the weak lensing signal of a reasonable
number of haloes. The main observational limitation is perhaps the
precise estimation of the turnaround radius, $r_{\rm ta}$, of superimposed
haloes. However, the loss of a few percents relative accuracy
in the determination of $r_{\rm ta}$ (or asphericity) can be compensated for
by stacking about $600$ haloes.
Wide field surveys such as the ongoing CFHTLS accompanied by X-ray observations
can provide the required statistics for a successful detection of caustics.
The number of haloes required to be superimposed will be lowered by a
further factor of 3 for future space-based experiments like SNAP or DUNE.




\end{document}